\documentclass[a4paper,12pt]{article}
\usepackage{feynmp-auto,expdlist}
\usepackage{amsmath, amsfonts}
\usepackage{graphicx,arydshln}
\usepackage{enumerate}
\usepackage{hyperref}
\usepackage{latexsym}
\usepackage{dsfont}
\usepackage{hepnicenames}
\usepackage{enumerate}
\usepackage{soul}
\usepackage[normalem]{ulem}
\usepackage{comment}

\newcommand{\footnoten}[1]{}

\usepackage[font={small},textfont={it}]{caption} 

\usepackage{units}

\usepackage{mathrsfs,graphicx,rotating,amsmath,amsfonts,mathtools,booktabs,amssymb,wasysym}
\usepackage{hyperref}\usepackage{slashed}
\usepackage[nosort]{cite}
\usepackage[table,xcdraw,dvipsnames]{xcolor}
\usepackage{bm}
\usepackage{multirow,multicol}
\hypersetup{colorlinks,bookmarksopen,bookmarksnumbered,
	linkcolor=blus,pdfstartview=FitH,urlcolor=rossos,citecolor=verde}
\allowdisplaybreaks

\newcommand{\myfootnote}[1]{}
\newcommand{\myomit}[1]{{\color{gray}#1}}
\renewcommand{\myomit}[1]{}

\renewcommand{\[}{\left[}

\def\Lag{\mathscr{L}}

\newcommand{\mio}[1]{}

\newcommand{\med}[1]{\langle #1\rangle}

\def\bpm{\begin{pmatrix}}
	\def\epm{\end{pmatrix}}

\usepackage{mathrsfs}

\newcommand{\fig}[1]{~\ref{fig:#1}}
\newcommand{\sfrac}[2]{#1/#2}

\allowdisplaybreaks
\usepackage{multicol}
\usepackage{color}
\definecolor{rosso}{cmyk}{0,1,1,0.4}
\definecolor{rossos}{cmyk}{0,1,1,0.55}
\definecolor{rossoc}{cmyk}{0,1,1,0.2}
\definecolor{blu}{cmyk}{1,1,0,0.3}
\definecolor{blus}{cmyk}{1,1,0,0.6}
\definecolor{bluc}{cmyk}{1,1,0,0.1}
\definecolor{verde}{cmyk}{0.92,0,0.59,0.25}
\definecolor{verdec}{cmyk}{0.92,0,0.59,0.15}
\definecolor{verdes}{cmyk}{0.92,0,0.59,0.4}

\newcommand{\eq}[1]{~{\rm (\ref{eq:#1})}}

\newcommand{\keV}{\,{\rm keV}}

\newcommand{\GeV}{\,{\rm GeV}}

\newcommand{\cm}{\,{\rm cm}}

\def\circa#1{\,\raise.3ex\hbox{$#1$\kern-.75em\lower1ex\hbox{$\sim$}}\,}

\newcommand{\beq}{\begin{equation}}
\newcommand{\eeq}{\end{equation}}

\newcommand{\bea}{\begin{eqnarray}}
\newcommand{\eea}{\end{eqnarray}}
\newcommand{\be}{\begin{equation}}
\newcommand{\ee}{\end{equation}}
\font\tenrsfs=rsfs10 at 12pt
\font\sevenrsfs=rsfs7
\font\fiversfs=rsfs5
\newfam\rsfsfam
\textfont\rsfsfam=\tenrsfs
\scriptfont\rsfsfam=\sevenrsfs
\scriptscriptfont\rsfsfam=\fiversfs

\newsavebox\MBox

\renewenvironment{thebibliography}[1]
{\begin{multicols}{2}[\section*{\refname}]%
		\@mkboth{\MakeUppercase\refname}{\MakeUppercase\refname}%
		\list{\@biblabel{\@arabic\c@enumiv}}%
		{\settowidth\labelwidth{\@biblabel{#1}}%
			\leftmargin\labelwidth
			\advance\leftmargin\labelsep
			\@openbib@code
			\usecounter{enumiv}%
			\let\p@enumiv\@empty
			\renewcommand\theenumiv{\@arabic\c@enumiv}}%
		\sloppy
		\clubpenalty4000
		\@clubpenalty \clubpenalty
		\widowpenalty4000%
		\sfcode`\.\@m}
	{\def\@noitemerr
		{\@latex@warning{Empty `thebibliography' environment}}%
		\endlist\end{multicols}}

\newcommand{\eV}{\,{\rm eV}}

\renewcommand{\L}\Lag

\def\circa#1{\,\raise.3ex\hbox{$#1$\kern-.75em\lower1ex\hbox{$\sim$}}\,}
\makeatletter

\font\ital=cmu10

\def\hhref#1{\href{http://arxiv.org/abs/#1}{arXiv:#1}}
\usepackage{xstring}
\newcommand{\hhrefq}[1]{\IfSubStr{#1}{:}{\href{http://inspirehep.net/search?ln=en&ln=en&p=#1&of=hb&action_search=Search&sf=&so=d&rm=&rg=25&sc=0}{InSpire:#1}}{\hhref{#1}}}

\def\art{\@ifnextchar[{\eart}{\oart}}
\def\eart[#1]#2#3#4#5#6{{\rm #2}, {\em #3 \bf #4} {\rm (#6) #5} ({\em #1})}
\def\article{\@ifnextchar[{\earticle}{\oarticle}}
\def\oarticle#1#2#3#4#5#6{{\rm #1}, {\ital `#6'}, {\rm #2 #3 (#5) #4}}
\def\earticle[#1]#2#3#4#5#6#7{{\rm #2}, {\ital `#7'}, {\rm #3 #4 (#6) #5}  [\hhrefq{#1}]}
\def\hepart[#1]#2{{\rm #2, \sl#1}}
\def\heparticle[#1]#2#3{#2, {\ital `#3'} [\hhrefq{#1}]}
\newcommand{\doi}[1]{\href{http://dx.doi.org/#1}{[link]}}

\newcommand{\hhrefqq}[1]{\IfBeginWith{#1}{10.}{\href{https://doi.org/#1}{doi:#1}}{\hhrefq{#1}}}
\def\earticle[#1]#2#3#4#5#6#7{{\rm #2}, {\ital `#7'}, {\rm #3 #4 (#6) #5}  [\hhrefqq{#1}]}

\renewenvironment{thebibliography}[1]
{\begin{multicols}{2}[\section*{\refname}]%
		\@mkboth{\MakeUppercase\refname}{\MakeUppercase\refname}%
		\list{\@biblabel{\@arabic\c@enumiv}}%
		{\settowidth\labelwidth{\@biblabel{#1}}%
			\leftmargin\labelwidth
			\advance\leftmargin\labelsep
			\@openbib@code
			\usecounter{enumiv}%
			\let\p@enumiv\@empty
			\renewcommand\theenumiv{\@arabic\c@enumiv}}%
		\sloppy
		\clubpenalty4000
		\@clubpenalty \clubpenalty
		\widowpenalty4000%
		\sfcode`\.\@m}
	{\def\@noitemerr
		{\@latex@warning{Empty `thebibliography' environment}}%
		\endlist\end{multicols}}

\newcounter{alphaequation}[equation]
\def\thealphaequation{\theequation\hbox to
	0.6em{\hfil\alph{alphaequation}\hfil}}
\def\eqnsystem#1{
	\def\@eqnnum{{\rm (\thealphaequation)}}
	\def\@@eqncr{\let\@tempa\relax \ifcase\@eqcnt \def\@tempa{& & &} \or
		\def\@tempa{& &}\or \def\@tempa{&}\fi\@tempa
		\if@eqnsw\@eqnnum\refstepcounter{alphaequation}\fi
		\global\@eqnswtrue\global\@eqcnt=0\cr}
	\refstepcounter{equation} \let\@currentlabel\theequation \def\@tempb{#1}
	\ifx\@tempb\empty\else\label{#1}\fi
	\refstepcounter{alphaequation}
	\let\@currentlabel\thealphaequation
	\global\@eqnswtrue\global\@eqcnt=0 \tabskip\@centering\let\\=\@eqncr
	$$\halign to \displaywidth\bgroup \@eqnsel\hskip\@centering
	$\displaystyle\tabskip\z@{##}$&\global\@eqcnt\@ne
	\hskip2\arraycolsep\hfil${##}$\hfil& \global\@eqcnt\tw@\hskip2\arraycolsep
	$\displaystyle\tabskip\z@{##}$\hfil
	\tabskip\@centering&\llap{##}\tabskip\z@\cr}
\def\endeqnsystem{\@@eqncr\egroup$$\global\@ignoretrue} \makeatother

\definecolor{Gray}{gray}{0.95}

\def\bal#1\eal{\begin{align}#1\end{align}}

\begin{document}
\thispagestyle{empty}

\begin{center}
{\LARGE \bf \color{rossos} 
Probing the Dark Matter density\\[0.3ex]
 with gravitational waves from\\[1ex]
super-massive binary black holes}

\bigskip\bigskip

{\bf\large Anish Ghoshal$^a$, Alessandro Strumia$^b$}  \\[5mm]
{$^a$ \em Institute of Theoretical Physics, Faculty of Physics, University of Warsaw, Poland }\\[1ex]
{$^b$ \em Dipartimento di Fisica, Universit{\`a} di Pisa, Italia}\\[4ex]

{\bf\color{blus} Abstract}
\begin{quote}
\large 
Supermassive black hole binaries source gravitational waves measured by Pulsar Timing Arrays.
The frequency spectrum of this stochastic background is predicted more precisely than its amplitude.
We argue that Dark Matter friction can suppress the spectrum around nHz frequencies, where it is measured,
allowing to derive robust and significant bounds on the Dark Matter density,
which, in turn, controls indirect detection signals from galactic centers.
A precise spectrum of gravitational waves would translate in a tomography 
of the DM density profile, potentially probing DM particle-physics effects that induce a characteristic DM density profile,
such as DM annihilations or de Broglie wavelength.
\end{quote}
\end{center}

\bigskip
\setcounter{tocdepth}{1}
\tableofcontents
\newpage
\normalsize

\section{Introduction}
Pulsar Timing Array (PTA) experiments report the observation of gravitational waves~\cite{NanoGrav,PPTA,EPTA,CPTA}.
This discovery is made by studying correlated timing distortions of the sufficiently stable millisecond rhythm of successive light pulses emitted by pulsars. 
These timing distortions seem to exhibit the angular dependence expected for an isotropic background of spin 2 gravitational waves~\cite{NanoGrav,EPTA}, 
distinguishing them from spin 0 or spin 1 waves and other effects in the way
computed by Hellings and Down~\cite{Hellings:1983fr}.

\smallskip

PTA observe gravitational waves $h_c(f)$ at low frequencies $f\sim (0.1-1)\,{\rm nHz}$
corresponding to the inverse of the observation time~\cite{NanoGrav,PPTA,EPTA,CPTA}.
Bounds on the gravitational wave spectrum have been derived at even lower frequencies~\cite{2304.13042}.

Potential cosmic sources could produce gravitational waves in the nHz frequency range~\cite{Moore:2021ibq, Chang:2019mza,Neronov:2020qrl,Brandenburg:2021tmp, 2202.01131,Freese:2023fcr,Ghoshal:2023sfa,Fu:2023nrn}.
However, despite hopes that these gravitational waves could be generated by new physics phenomena, 
the observed amplitude and spectral slope are approximately consistent with the astrophysical background. 
The astrophysical background is expected to be primarily generated by binary supermassive black holes (SMBHs) with masses
 $M_{1,2}$ in the range $ \sim 10^{8-9}M_\odot$~\cite{1211.5377,1404.5183,1503.02662,1508.03024,1612.02817,1702.02180,1709.06095,1811.08826,2105.04559,2301.13854,2302.00702}.
The expected amplitude of the astrophysical background is subject to an order of magnitude uncertainty due to several factors. Firstly, the density of supermassive black hole (SMBH) sources is uncertain. Additionally, the properties of these sources, such as the distribution of masses $M_{1,2}$, eccentricity, and redshift, are also uncertain. 
Moreover, the possible contribution of nearby sources  further adds to the uncertainty in the amplitude estimation.
On the other hand, the expectation for the frequency spectrum of the astrophysical background is more precise. 
This precision arises from the characteristic behaviour of non-relativistic binary systems that predominantly lose energy through gravitational wave emission. Specifically, the frequency spectrum emitted by such systems follows a power-law relationship, with $h_c(f) \propto f^{-2/3}$.


\smallskip

Interestingly, in the observed frequency range, a change in the spectral slope of emitted gravitational waves can occur due to the interaction of supermassive black holes with ambient matter and Dark Matter (DM). As SMBHs move through their surroundings, they experience friction with both the ambient matter and DM, resulting in energy loss that competes with gravitational wave emission.
This effect of DM friction has been previously considered in the context of binary pulsars
(leading to constraints on the DM density $\rho_{\rm DM}\circa{<} 10^5\GeV/\cm^3$~\cite{1512.01236}),
intermediate mass BH~\cite{1404.7140,1408.3534,2002.12811,2108.04154,2210.01133},
and binary solar-mass BH (two anomalies
are interpreted as the possible observation
of $\rho_{\rm DM} \sim 10^{11-13}\GeV/\cm^3$ in~\cite{2212.05664}).
We find that a better probe is now offered by super-massive BH~\cite{1211.5377,1404.5183,1503.02662,1508.03024,1612.02817,1702.02180,1709.06095,1811.08826,2105.04559,2301.13854,2302.00702}, 
as gravitational waves observe
their dynamics while they are non-relativistic, in the relevant frequency range for this effect.

\medskip

This paper is organised as follows: the effect of DM friction 
is computed in section~\ref{DMGW}, and compared to existing data in section~\ref{data}. Conclusions are given in section~\ref{concl}.

\section{Dark Matter influence on gravitational waves}\label{DMGW}
We consider two black holes with masses $M_1$ and $M_2$ orbiting at distance $r$ with angular velocity $\omega$ and velocity $v$.
Energy losses due to gravitational radiation are expected to dominate at smaller $r$ (corresponding to larger $\omega$),
while energy losses due to matter and DM friction are expected to dominate at larger $r$ (smaller $\omega$).
The spectrum of emitted gravity waves changes slope at the critical radius $r_{\rm cr}$ and critical frequency $f_{\rm cr}$
where the two mechanisms contribute in a comparable way. 
At the nHz frequencies observed by PTA, ultra-massive BH binaries orbit at distances
much larger than their Schwarzschild radius $R_{\rm Sch} =2GM\approx 10^{-5}\,{\rm pc} \, M_8$ where $M_8\equiv  M/10^8\,M_{\rm sun}$.
So their motion is in the non-relativistic Newtonian regime $v\ll 1$.
Assuming, for simplicity, circular motion, the radial acceleration is
\beq  \label{eq:fisica1}\frac{v^2}{r}= \omega^2r = \frac{G(M_1+M_2)}{ r^2}\eeq
corresponding to $v=\omega r\sim \sqrt{R_{\rm Sch}/r} \ll 1$ for $M_1\sim M_2$.
Here $G$ is the Newton constant, and we use natural units.
The emitted gravitational waves have frequency $f_s= \omega/\pi $ around the source,
with no factor of 2 because gravitational waves have spin 2.
The observed frequency is $f=f_s/(1+z)$ if the source is located at redshift $z$.
So the  frequency range $f\sim{\rm nHz}$ observed by PTA is emitted when the orbital distance is
$r/R_{\rm Sch}\approx 2000/M_8^{2/3}$,
and while SMBHs orbit with a non-relativistic velocity comparable to the velocity of ambient galactic DM particles.
This makes gravitational waves from SMBH a better probe of the DM density than solar-mass BH, 
whose gravitational waves are currently observed in the final stage when they are relativistic.

\smallskip

\begin{figure}[t]
\begin{center}
$$\includegraphics[width=\textwidth]{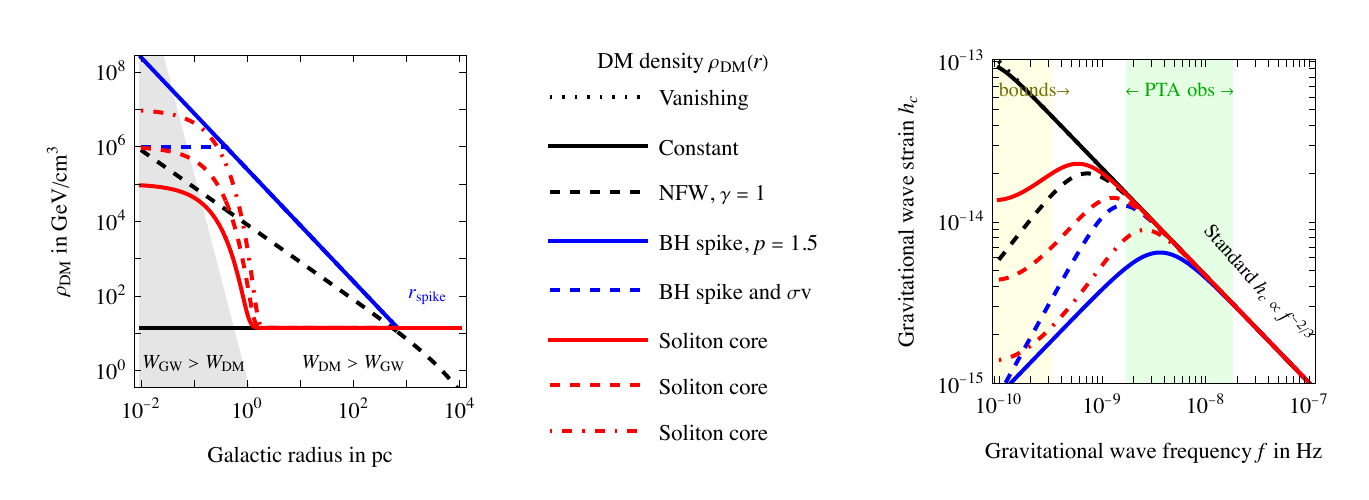}$$
\caption{\label{fig:spectrumS} {\bfseries Left}: illustrative samples of possible DM density profiles around binary super-massive black holes.
{\bfseries Right:} consequent spectra of emitted gravitational waves, taking into account dynamical friction due to DM with $\wp \ell=5$.
Observations~\cite{NanoGrav,PPTA,EPTA,CPTA} (bounds~\cite{2304.13042}) exist in the frequency range shaded in green (yellow).
We here assume the BH masses $M=10^{8.5}M_\odot$ expected to dominate the gravitational wave spectrum.
Energy losses due to DM friction are subdominant compared to GW energy losses in the shaded region of the left plot,
leading to suppressed deviations at higher frequency in the right plot.} 
\end{center}
\end{figure}

We consider the two following mechanisms for energy loss:
\begin{itemize}
\item The power radiated via gravitational waves is $W_{\rm GW}\sim v^{10}/G$ or more precisely~\cite{Maggiore}
\beq W_{\rm GW}=\frac{32}{5} G \mu^2 \omega^6 r^4,\qquad \eeq
where $\mu = M_1 M_2/(M_1+M_2)$ is the reduced mass, so binaries with larger $M_1\sim M_2$ dominate $W_{\rm GW}$.

\item The energy loss due to DM swept by a BH is estimated as
\beq \label{eq:WDMnaive} W_{\rm DM} \sim \pi b^2\,  \delta v\, \rho_{\rm DM}\qquad
\hbox{where} \qquad b \sim R_{\rm Sch} /\delta v\eeq 
is the impact parameter below which the BH captures DM particles with relative velocity $\delta v$ at large distance.
The impact parameter $b$ is larger than the BH radius $R_{\rm Sch}$ because gravitational attraction provides the classical Sommerfeld enhancement in eq.\eq{WDMnaive}.
This energy loss is more precisely computed as the `dynamical friction' due to the gravity of DM attracted by the BH~\cite{Chandrasekhar:1943ys,1802.03739,2212.05664}
\beq  \label{eq:dynfr}W_{\rm DM} = \frac{4\pi G^2 \mu^2 \rho_{\rm DM}}{v} \wp \ell.\eeq
Here $\ell \sim 10$ is a logarithm that arises because gravity is a long-range force,
and $\wp \sim 1/2$ is the fraction of DM slower than $v$.
Both factors have mild astrophysical uncertainties.
\end{itemize}
The energy loss due to DM friction can dominate, giving $W_{\rm DM} > W_{\rm GW}$, 
when $r$ is larger than some critical value $r >r_{\rm cr}$ corresponding to a low $\omega < \omega_{\rm cr}$ 
\beq \label{eq:DMr} 
 \frac{r_{\rm cr}}{R_{\rm Sch}}\approx\frac{1}{G^{6/11} M^{4/11} \rho_{\rm DM}^{2/11}} 
 ,\qquad
\omega_{\rm cr} \approx \frac{\rho_{\rm DM}^{3/11}}{G^{2/11} M^{5/11}}\approx \frac{0.23\,{\rm nHz}}{M_8^{5/11}} \left(\frac{\rho_{\rm DM}(r_{\rm cr})}{0.4\GeV/\cm^3}\right)^{2/11}.
\eeq
The grey shading in fig.\fig{spectrumS}a illustrates such critical boundary.
We omitted order one factors in the analytic expression, but we include them in the numerical computations
performed assuming $M_1=M_2=M$, $\wp=1/2$, $\ell=10$.
Eq.\eq{DMr} shows a potentially interesting value of $\omega_{\rm cr}$ even assuming
as reference value the DM density around the Sun, $\rho_{\rm DM}^\odot\approx  0.4 \GeV/\cm^3$.
As discussed later, this likely is an underestimate, 
as the DM density $\rho_{\rm DM}(r)$ is expected to be larger around galactic centers,
and an extra DM density spike is expected around black holes.

\subsubsection*{The spectrum of gravitational waves}
The energy spectrum of emitted gravitational waves
is simply computed taking into account that
circular rotation with angular velocity $\omega$ produces a mono-chromatic gravitational wave with frequency $f_s= \omega/\pi$ at source,
so that $\sfrac{dE_{\rm GW}}{d\omega} = \sfrac{W_{\rm GW}}{d\omega/dt}$.
The secular change in $\omega$ due to all energy losses, $\dot\omega$, is computed 
by imposing conservation of energy 
$\dot E =- W_{\rm GW}- W_{\rm DM}$, where
$E = \mu v^2/2 - GM_1 M_2/r$,
taking into account that $\omega$ and $r$ are related as in eq.\eq{fisica1}.
The total gravitational wave spectrum simplifies in the limits where either GW or DM dominate the energy loss:
\beq \label{eq:EGW}
\frac{dE_{\rm GW}}{d\omega} =
\left\{\begin{array}{ll}\displaystyle
\frac{M_1 M_2 G^{2/3}}{3(M_1+M_2)^{1/3} }\omega^{-1/3} & \hbox{GW-dominated,}\\[1ex] \displaystyle
\frac{8 G^{4/3} M_1 M_2 (M_1+M_2)^{4/3} \omega^{10/3}}{15\pi \wp \ell\, \rho_{\rm DM}(G^{\rm 1/3}(M_1+M_2)^{1/3}/\omega^{2/3})}&
\hbox{DM-dominated.}
\end{array}\right.
\eeq
The above contribution of one SMBH source must be integrated over all SMBHs, obtaining
the cosmological average GW energy density $d\rho_{\rm GW}/df$, usually written in terms of the  characteristic strain spectrum $h_c(f)$ as
(see e.g.~\cite{1408.0740,astro-ph/0108028})\footnote{Dividing the GW energy density 
by the critical density $\rho_{\rm cr} = 3H_0^2/8\pi G$
one also defines the dimensionless
$\Omega_{\rm GW} \equiv d\rho_{\rm GW}/d\ln f/\rho_{\rm cr}$.
GW-dominated circular binaries produce $h_c \propto f^{-2/3}$ corresponding to $\Omega_{\rm GW}  \propto f^{2/3} $. }
\beq h_c^2(f) = \frac{4G}{\pi  f^2} \frac{d\rho_{\rm GW}}{d\ln f}=
 \frac{4G}{\pi f^2} 
\int \frac{dz}{1+z}\int dX\frac{dN_s}{dz \, d X} \frac{dE_{\rm GW}}{d\ln f_s}\eeq
where $f=f_s/(1+z)$ and
$X$ collectively denotes all parameters that describe the SMBH sources (the masses $M_{1,2}$,
the ellipticity...) at red-shift $z$.
Models and data indicate a source distribution $dN_s/dz\,dX$ such that
the integrals are dominated around red-shift $z\sim 0-0.3$ and for masses $\max M_{1,2}\sim 10^{8-9}M_\odot$~\cite{2105.04559}.

\subsubsection*{The expected DM density profile}
Eq.\eq{EGW} shows that gravitational waves provide a tomography of a generic (dark) matter density profile:
each density profile $\rho_{\rm DM}(r)$ translates into a spectrum $h_c(f)$ of gravitational waves,
with motion at larger $r$ testing lower values of the frequency $f$.
In the regime where  the DM density is so small that $W_{\rm DM}\ll W_{\rm GW}$, 
the usual $dE_{\rm GW}/d\omega \propto \omega^{-1/3}$ only receives small corrections due to DM friction.
Significant corrections arise in the regime where $W_{\rm DM}\gg W_{\rm GW}$.
A constant $\rho_{\rm DM}$ leads to a sharply different spectral slope,
$dE_{\rm GW}/d\omega \propto \omega^{10/3}$.
A similar slope arises assuming a NFW density profile~\cite{NFW}, illustrated by dashed black curve in fig.\fig{spectrumS}.

\smallskip

Rotation curves allow to infer the value of the DM density  in outer regions of galaxies.
For example $\rho_{\rm DM}^\odot \approx 0.4\GeV/\cm^3$ at the Sun position in the Milky Way, $r_\odot \sim 8\,{\rm kpc}$.
Assuming that DM behaves as perfect dust, 
its density is expected to be enhanced around the Galactic Centers, and especially around a SMBH.
We define $r_{\rm in}$  as the `influence radius' around a SMBH, 
namely the radius within which the total mass
is dominated by the SMBH mass: $r_{\rm in}\approx (3M/2\pi \rho_{\rm DM})^{1/3}  \sim {\rm kpc}$.
A BH spike is expected to be present at  $r< r_{\rm spike}\approx 0.2 r_{\rm in}$
(an example is shown in fig.\fig{spectrumS}a).
Defining $ \rho_{{\rm DM}\,{\rm GC}}$ as the DM density at $r=r_{\rm spike}$,
computations and simulations suggest it is enhanced as
\beq \label{eq:rhoDMGC}
 \rho_{{\rm DM}\,{\rm GC}}\approx \rho_{\rm DM}^\odot (r_\odot/r_{\rm spike})^{p'}\eeq
with $p'=1$ for the NFW profile~\cite{NFW}, and similar values of $p'$ for other widely used  profiles.
We take the Milky Way as a concrete typical example.
The possible BH spike on top of the galactic density
is estimated as (see e.g.~\cite{astro-ph/9906391,astro-ph/9407005,2212.05664,2303.12107}) 
\beq \label{eq:DMspike}\rho_{\rm DM}(r)\approx  \rho_{{\rm DM}\,{\rm GC}} 
(r_{\rm spike}/r)^p \qquad \hbox{at $r<r_{\rm spike}$}
\eeq
with power $p\approx 1.5-2.5$.
According to computations that assume adiabatic growth and that are not tested by simulations,
a larger $p'$ is expected to increase the spike power index as $p \approx (9-2p')/(4-p')$~\cite{astro-ph/9906391,2303.12107}.
As $r_{\rm in}$ is typically  bigger than the binary orbital radius in the observed gravitational frequency range, 
we can neglect the DM contribution to the gravitational force acting among SMBHs,
and a simple parameterisation of the form of eq.\eq{DMspike} is general enough to compute DM friction.
Eq.\eq{DMspike}  can describe in a unified way the galactic DM density profile for $p\approx 1$ (in case no BH DM spike exists)
and a BH DM spike (for $p\circa{>}1.5$, in case it exists).
In the following we also conservatively consider the case $p=0$, 
corresponding to no DM density enhancement around the Galactic Center nor around the SMBH,
to cover the possibility that the galactic merging event that lead to a SMBH binary might have disrupted the DM distribution.
We however expect that the standard galactic density profile formed again after the galactic merging,
and that pre-existing DM spikes survived during the galaxy merging, thanks to their gravitational binding,
especially if $M_1\ll M_2$.

\smallskip

Inserting the BH profile of eq.\eq{DMspike} in eq.\eq{DMr}  for the critical frequency allows to express $\omega_{\rm cr}$
in terms of galactic density $ \rho_{{\rm DM}\,{\rm GC}}$ around the galactic centers outside the
influence radius of the SMBHs:
\beq \label{eq:fcrp}
\omega_{\rm cr}\approx [G^{2+p} M^5 \rho_{{\rm DM}\,{\rm GC}}^{p-3}]^{1/(2p-11)}\approx
\frac{13\,{\rm nHz}}{M_8^{5/7}}\left(\frac{ \rho_{{\rm DM}\,{\rm GC}}}{0.4\GeV/\cm^3}\right)^{3/16}
\qquad\hbox{for power $p=1.5$}.
\eeq
We again use a conservative reference value for $ \rho_{{\rm DM}\,{\rm GC}}$.
Even so, the resulting critical frequency $f_{\rm cr}= \omega_{\rm cr}/\pi(1+z)$ is around
the range where gravitational waves have been measured by PTA~\cite{NanoGrav,PPTA,EPTA,CPTA},
showing that DM friction is significant.
Inserting eq.\eq{DMspike} in eq.\eq{EGW}  for the DM-dominated gravitational wave spectrum gives
\beq\label{eq:dEGWp} \frac{dE_{\rm GW}}{d\omega} \approx \frac{G^{(4+p)/3} M^{10/3}}{ \rho^{1-p/3}_{{\rm DM}\,{\rm GC}}} 
 \omega^{2(5-p)/3}  .\eeq
The spectral index in  eq.\eq{dEGWp} significantly differs from the index in the GW-dominated limit:
the same index only arises if the DM density grows at small $r$ as $p= 11/2$, outside the range suggested by structure formation. 
Moving to plausible values of $p$,
the dashed black curve in fig.\fig{spectrumS}b illustrates the frequency spectrum
resulting from a NFW DM density profile,
and the blue curve in fig.\fig{spectrumS} does the same for a BH DM spike on top of NFW.
The expected DM friction effect is significant.


\medskip

Having discussed the expectation of dust DM, we next discuss 
how specific DM particle physics effects can induce characteristic profiles $\rho_{\rm DM}(r)$ that would
be observable as characteristic features of $dE_{\rm GW}/d\omega$ or 
(after averaging over SMBH sources, and loosing some information) of $h_c(f)$.
\begin{itemize}
\item If DM is particles with mass $m$ that annihilate with cross section $\med{\sigma v}$,
the enhancement  of eq.\eq{DMspike} gets limited as
$\rho_{\rm DM}(r) \circa{<} m/\med{\sigma v} \tau_{\rm BH}$
in a BH with age $\tau_{\rm BH} \sim 10^{10}\,{\rm yr}$ (see e.g.~\cite{2303.12107}).
Testable values of the saturation density arise for $m\circa{<}\GeV$.
The blue dashed curve in fig.\fig{spectrumS} illustrates how this DM density profile translates in a frequency spectrum.

\item A variant of the above scenario is DM with peculiar kinematics such that the kinetic energy 
acquired from the gravitational potential of the BH opening a new annihilation or scattering 
 at some distance from the BH, with implications on the DM density profile~\cite{2211.05643}.

\item If DM is ultra-light free bosons with mass $m\sim 10^{-21}\eV$,
a soliton-like configuration can form around a BH, in the region where gravity is dominated by the BH mass~\cite{1805.00122,1905.11745}.
Particle physics fixes its properties as later discussed around eq.\eq{soliton}.
Focusing on the soliton,
its cored density profile leads to the spectrum of gravitational waves illustrated by the red curves in fig.\fig{spectrumS}
for different soliton densities.

\item Another possibility is that (a part of) DM is fermions with small enough mass $m\circa{<} \keV$ that  
Fermi-Dirac repulsion starts affecting its density distribution when $\rho_{\rm DM}$ reaches $\sim m^4 v_{\rm DM}^3$
in the inner part of the BH spike, provided that accretion can be slow enough.
\end{itemize}

\subsubsection*{Astrophysical uncertainties}
\begin{figure}[t]
\begin{center}
$$\includegraphics[width=0.6\textwidth]{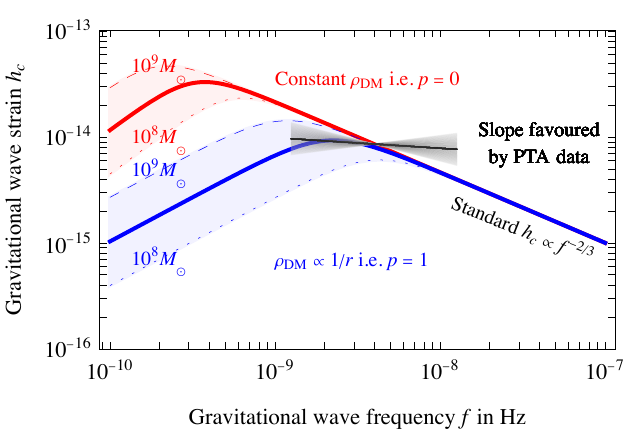}$$
\caption{\label{fig:spectrum} Softening of the gravitational wave spectrum at low frequency due to DM friction.
We assume a circular binary of SMBH with the equal indicated masses located
at redshift $z=0.15$, with DM density around the influence radius of the SMBH assumed to be
$ \rho_{{\rm DM}\,{\rm GC}}=10^3\GeV/\cm^3$.
 The DM density is assumed to scale as $1/r^p$ at smaller distances, with $p=0$ for the red curves and $p=1$ for the blue curves.
 The grey band represents the range of spectral slopes favoured by PTA observations~\cite{NanoGrav,PPTA,EPTA,CPTA} at $\pm1\sigma$.
For simplicity we omit the uncertainty in the overall normalization of gravitational waves,
as we only use the information contained in their spectral slope.} 
\end{center}
\end{figure}

The astrophysical prediction for the overall normalization of $h_c(f)$ depends on the number density of SMBH sources and
is significantly uncertain, so we will only rely on the spectral shape $\beta=d\ln h_c/d\ln f$.
The red-shift correction is expected to be mild, and the main uncertainty is
expected to come from the distribution of SMBH masses $M_{1,2}$.
Fig.\fig{spectrum} illustrates the frequency spectra produced by a constant $\rho_{\rm DM}(r)$ (red)
and by $\rho_{\rm DM}(r)\propto 1/r$ (blue) for different values of $M_1=M_2 = M  = 10^8M_\odot$
(dotted, on the lower side of the expected SMBH range) and $10^9M_\odot$ (dashed, upper side).
We see that the features in $\rho_{\rm DM}$  get partially washed out when averaging over the distribution of SMBH masses.

Furthermore, two additional effects are expected to correct the spectrum $h_c(f)$ in a way qualitatively similar to dynamical friction due to DM:
\begin{itemize}
\item Dynamical friction due to baryonic matter, such as stars and gas, that possibly formed a disk~\cite{1211.5377,1404.5183,1503.02662,1508.03024,1612.02817,1702.02180,1709.06095,1811.08826,2105.04559,2301.13854,2302.00702}.
While the cosmological density of DM is about 5 times larger than the matter density, 
the situation around the center of merged galaxies is uncertain and possibly 
baryonic matter is more dense that DM.

\item Ellipticity of the SMBH binary
makes the gravitational waves emitted during one orbit non monochromatic, 
enhancing $h_c(f)$ at higher $f$ as the gravitational acceleration grows at smaller distances~\cite{1404.5183}.
The ellipticity distribution of the relevant SMBH sources is uncertain.
Energy losses tend to reduce the ellipticity, and ellipticity makes dynamical friction mildly more relevant~\cite{1404.5183}.
\end{itemize}
These extra uncertain astrophysical effects qualitatively contribute similarly to DM dynamical friction, 
complicating the possibility of reconstructing the DM density profile from gravitational waves.
For the moment we conservatively use current data to set a robust upper bound on the DM density, that turns out to be numerically interesting.

\begin{figure}[t]
\begin{center}
$$\includegraphics[width=0.8\textwidth]{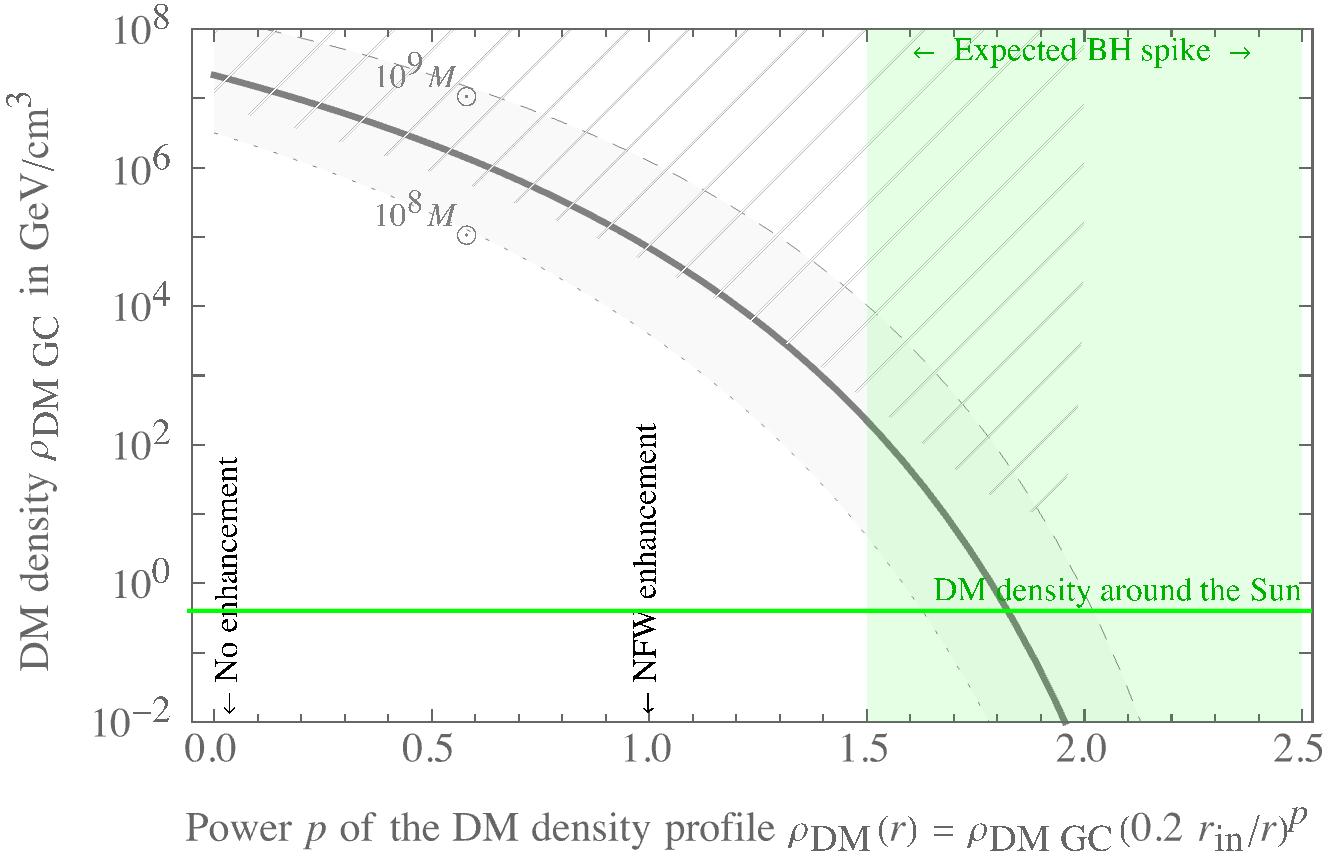}$$
\caption{\label{fig:BHspikeDMdensity} 
Parameterizing the DM density profile as
$ \rho_{{\rm DM}}(r)= \rho_{{\rm DM}\,{\rm GC}}(0.2 r_{\rm in}/r)^p$
where $r_{\rm in} \sim \,{\rm kpc}$ is the influence radius of the SMBH,
we show the gravitational wave bound on $p$ and on the DM density around the BH, $\rho_{{\rm DM}\,{\rm GC}}$.
The NFW profile has $p=1$; a BH DM spike has $p\approx 1.5-2.5$.
The hatched region above the grey band is excluded. 
The band is the analytic bound $f_{\rm cr}<1/(10\,{\rm yr})$ of eq.\eq{boundsimpl} 
computed for BH masses $M=10^{8,8.5,9}M_\odot$ and red-shift $z=0.15$.
The hatching is the more precise bound of eq.\eq{boundslope} for $M=10^{8.5}M_\odot$.} 
\end{center}
\end{figure}

\section{Comparison with Pulsar Time Array data}\label{data}
Several authors studied the gravitational wave background from SMBH mergers, including the range probed by PTA~\cite{1901.06785,1709.06501,astro-ph/0609377,1006.0730,1211.5375,1307.2600}.
As the range covered by PTA is limited and uncertainties are significant,
PTA analyses parameterize the strain spectrum as a simple power law
$h_c(f)=A_{\rm GW} (f/f_{\rm PTA})^\beta$ where $f_{\rm PTA}$ is an arbitrary reference frequency,
and measure the power spectrum of temporal correlations 
$P_{\rm GW}=A_{\rm GW}^2(f_{\rm PTA}/f)^{\gamma_{\rm GW}}/12\pi^2$, 
fitting it in terms of $A_{\rm GW}$ and $\gamma_{\rm GW}=3-2\beta$. 

While the complex astrophysical environments of SMBH makes the emitted gravitational waves uncertain, 
their spectral slope $\beta$ is robustly predicted in the limit where energy losses are dominated
by emission of gravitational radiation.
Whether gravitational waves are the dominant energy loss mechanism 
or not has been a topic of active discussion in the literature and is known as the ``final parsec problem''~\cite{Milosavljevic:2002ht} (see e.g.~\cite{1102.4855,1103.0272,1607.04284,1505.05480,1609.09383,1712.05810,2201.08646} for discussions on possible resolutions and~\cite{1503.02662,1612.02817} for tests via PTA measurements at frequencies above 1 nHz).
As discussed above
\begin{itemize}
\item[{\sc gw})] Circular binaries dominantly driven by gravitational waves give $\beta=-2/3$, corresponding to $\gamma_{\rm GW}=13/3$.
\item[{\sc dm})]  Circular binaries dominantly driven by dynamical friction due to DM with a density profile $\rho_{\rm DM}(r) \propto r^{-p}$
give $\beta= (7-2p)/6>-2/3$, corresponding to $\gamma_{\rm GW}=2(1+p)/3 < 13/3$.
\end{itemize}
The experimental situation can be summarised as follows.
Choosing $f_{\rm PTA}=1/(10\,{\rm yr})= 3.2\,{\rm nHz}$ in the middle of the measured frequency range
avoids a spurious correlation between $A_{\rm GW}$ and $\gamma_{\rm GW}$
that arises if $f_{\rm PTA}=1/{\rm yr}$ is employed.
{\sc NanoGrav}~\cite{NanoGrav} finds 
$\gamma_{\rm GW}\approx 3.2 \pm 0.6$ around the frequency  $1/(10\,{\rm yr})$, 
altought  earlier data preferred a higher $\gamma_{\rm GW}$.
EPTA finds 
$\gamma_{\rm GW}=4.0\pm 0.4$ around  the frequency  $1/{\rm yr}$ and
$\gamma_{\rm GW}=2.7 \pm 0.9$ around  the frequency  $1/(10\,{\rm yr})$.
PPTA finds $\gamma_{\rm GW}=3.9\pm 0.5$ using $1/{\rm yr}$ as reference frequency and cannot discriminate the spin of the wave signal~\cite{PPTA}.
CPTA cannot significantly constrain the spectral index~\cite{CPTA}.
Each collaboration performs multiple analyses, and the results from the various collaborations cannot be naively combined,
as their data-sets overlap.
Results of previous data releases 
have been combined by IPTA~\cite{IPTA}, finding $\gamma_{\rm GW}=3.8\pm 0.6$ at $f_{\rm PTA}=1/10\,{\rm yr}$.

\medskip

These spectral measurements disfavour gravitational waves emitted by SMBH with
DM-dominated energy losses, covering plausible values of the power $p$.
Thereby a simple  approximate analytic bound on the DM density is obtained by imposing that the critical frequency below
which DM friction dominates,
$f_{\rm cr}$ from eq.\eq{fcrp}, is below $ f_{\rm PTA}$:
\beq \label{eq:boundsimpl}
\rho_{{\rm DM}\,{\rm GC}} \circa{<} \left(\frac{f^{2p-11}_{\rm PTA}}{G^{2+p} M^5}\right)^{1/(p-3)} .
\eeq
Including order unity factors, fig.\fig{BHspikeDMdensity} shows such bound in the $(p, \rho_{{\rm DM}\,{\rm GC}})$ plane.
The dotted (dashed) curves assume $M=10^8 M_\odot$ ($M=10^9 M_\odot$); 
the continuous thick curve assumes $M = 10^{8.5}M_\odot$, the SMBH mass expected to dominate the GW spectrum.
We here used $f_{\rm PTA}=1/10\,{\rm yr}$.
This choice gets justified by the agreement with the same bound imposed in a more precise way,
as a constraint on the spectral index computed taking into account
both GW and DM energy losses.
Focusing on $M = 10^{8.5}M_\odot$, the hatching in fig.\fig{BHspikeDMdensity} covers the regions that violate 
\beq\label{eq:boundslope}
 \beta=\frac{d \ln h_c}{d\ln f}=\frac{3-\gamma_{\rm GW}}{2}< 0.5 \eeq
which might correspond to a $2-3\sigma$ bound, 
while waiting for a IPTA combination of the most recent data.
This bound matches well the analytic approximation of eq.\eq{boundsimpl}. 

\medskip

The bound on the DM density
is weakened by the fact that some PTA data~\cite{NanoGrav,PPTA,EPTA,CPTA,IPTA}
 suggest a $\approx 2\sigma$ hint for  $\gamma_{\rm GW}<13/3$.
This can be interpreted as a new-physics source of gravitational waves,
or more simply as SMBH sources affected by the dynamical friction of matter plus dark matter, 
such that $ f_{\rm cr} \approx f_{\rm PTA}$ and a spectral break will appear in future data.
If dark matter dominates, this hint would favour the regions around the exclusion boundary in fig.\fig{BHspikeDMdensity}.
This means that the DM density allowed (and perhaps favoured)  by gravitational wave data is
\beq \label{eq:allowed}
\rho_{{\rm DM}\,{\rm GC}}\circa{<}100\GeV/\cm^3\approx 2.6 M_\odot/{\rm pc}^3\qquad\hbox{for a BH density with power $p\approx 1.5$.}\eeq
The gravitational wave bound on the DM density is quantitatively significant,
as our understanding of galaxy formation suggests that
DM is expected to have around the center of galaxies like the Milky Way
a  density higher than $\rho_{\rm DM}^\odot \approx 0.4\GeV/\cm^3$, eq.\eq{rhoDMGC}.
The values of $ \rho_{{\rm DM}\,{\rm GC}}$ and $p$ 
allowed by gravitational waves according to fig.\fig{BHspikeDMdensity} or eq.\eq{allowed} 
appear compatible with expectations for typical galaxies.
One should keep in mind that gravitational waves probe atypical galaxies that underwent a major SMBH merger, that perhaps
smoothened the DM density distribution~\cite{astro-ph/0201376}.
Furthermore the formation of SMBH itself is not well understood.

\medskip

\subsubsection*{The case of ultra-light DM around SMBH}
Finally, let us focus on dynamical friction in the case of DM as a super-light boson with mass $m$.
Current gravitational wave data exclude a too dense ultra-light DM soliton formed around SMBH, in the region dominated by their gravity.
Indeed such a soliton  has a cored exponential density profile $e^{-r/r_{\rm soliton}}$
with  $r_{\rm soliton}\sim 1/m v_{\rm soliton}$ given by the DM wavelength.
As $v^2 = R_{\rm Sch}/r$ in the BH-dominated gravity (eq.\eq{fisica1}),
the soliton radius gets fixed as\cite{1905.11745}
\beq \label{eq:soliton}
r_{\rm soliton}\sim 1/m^2 R_{\rm Sch}, \qquad \hbox{with}\qquad v_{\rm soliton} \sim m R_{\rm Sch}.\eeq
We assume $v_{\rm soliton}\ll 1$ so that  super-radiance can be neglected~\cite{1004.3558}.
The soliton density is not fixed by fundamental physics; astrophysical simulations suggest that solitons form 
with total mass comparable to the galactic BH mass, resulting in $M_{\rm soliton}\sim 0.1 M$~\cite{1905.11745,1805.00122}.
We consider a $r_{\rm soliton}$ small enough that the DM solitons survive to the galactic merger,
and we approximate dynamical friction using eq.\eq{dynfr}; a dedicated computation would be appropriate and
could include larger effects from possible extra boson self-interactions~\cite{2007.03700}.
A soliton with the large expected density is excluded by data
as $r_{\rm soliton}\circa{<} r_{\rm cr}$ of eq.\eq{DMr} gives $m\circa{>} 10^{-20}\eV/M_8^{9/11}$
and $v^2_{\rm soliton}\circa{>}10^{-3} M_8^{2/11}$.
In this mass range BH accrete depleting the soliton in a time $t\sim 1/m v_{\rm soliton}^5$ 
less than $\tau_{\rm BH}\sim 10^{10}\,{\rm yr}$~\cite{1905.11745}.
A depleted soliton with density orders of magnitude below the expected one is compatible with data, 
as exemplified by the red curves in fig.\fig{spectrumS}.
Black holes binaries with lower masses $M_{1,2}\ll 10^8 M_\odot$ provide a small contribution to gravitational waves 
that can be strongly distorted by the intact soliton core: a dedicated study could be interesting.


\section{Discussions and conclusions}\label{concl}
We discussed the possibility of utilizing Pulsar Timing Array measurements of gravitational waves produced by supermassive black hole binaries to probe the Dark Matter density distribution $\rho_{\rm DM}(r)$ around such objects. The spectrum of gravitational waves emitted by supermassive black holes can be reliably predicted. The dynamical friction caused by the Dark Matter density can reduce the kinetic energy of supermassive black holes, competing with gravitational wave emission and resulting in a suppression of their frequency spectrum $h_c(f)$. 
This effect is particularly significant at lower frequencies, corresponding to larger orbital radii.
It can become dominant within the frequency band of gravitational waves observed by Pulsar Timing Arrays for realistic values of the Dark Matter density.

\smallskip

Current PTA data already allow to probe the DM density, and
favour a gravitational wave spectrum compatible with a mild dynamical friction effect.
We conservatively used current PTA data to set robust bounds on the DM density.
Assuming that DM is dust, astrophysics suggests a  higher DM density  around galactic centers,
and possibly an additional peculiar spike around SMBH.
The resulting bound on the DM density, obtained within this minimal  scenario, is shown in fig.\fig{BHspikeDMdensity}.
An analytical approximation for the constraint is provided  in eq.\eq{boundsimpl}.
Its quantitative significance is discussed in section~\ref{data}.

\smallskip

Furthermore, we discussed how gravity wave observations could provide a tomography of the DM density distribution, 
as each DM density profile $\rho_{\rm DM}(r)$ leads to a corresponding spectrum $h_c(f)$.
On top of astrophysical motivations, this could open new avenues for studying fundamental physics.
We discussed how fundamental particle-physics effects can give signatures in the DM density profile.
For instance, DM annihilations or Fermi repulsion could result in a saturation density, 
while ultra-light bosonic DM could give rise to a solitonic core. 
These distinct DM density profiles would have consequential effects on the gravitational wave spectrum, as depicted in figure \ref{fig:spectrumS}.
By examining the gravitational wave spectrum, we could gain insights into the underlying fundamental physics governing DM.

\smallskip

Observing such features would need precise data.
More precise observations of gravitational waves, 
in a frequency range extended up to $f\circa{<} \mu{\rm Hz}$, 
could be derived by looking at how star positions oscillate,
from surveys of stars such as {\sc Gaia} and the {\sc Theia} and possible future upgrade~\cite{1707.06239,2104.04778},
and from upcoming radio telescopes such FAST~\cite{1105.3794} and SKA~\cite{astro-ph/0409274}.
Future gravitational wave data from identified sources might allow to reduce the impact of astrophysical uncertainties.
This could possibly lead to multi-messenger probes of super-massive binary black holes:
connections with the  astrophysical flux of high-energy neutrinos probed by IceCube are suggested in~\cite{2210.11337}.

\smallskip

If gravitational waves will better determine the DM density profile,
this will allow to more precisely compute DM indirect signals from galactic centers, and to
improve the foreground modelling for future CMB experiments like Simon's observatory \cite{1808.07445} and CMB-S4 \cite{2203.07064} 
and future GW experiments. 
If future gravitational wave data were to imply that a non-zero dark matter density $\rho_{\rm DM}>0$ is 
present around SMBHs, 
this observation alone would disfavour alternatives to DM such as Modified Newtonian Dynamics (MOND~\cite{MOND}),
as MOND proposes modifications to the laws of gravity below a critical acceleration
orders of magnitude below the acceleration in eq.\eq{fisica1}.

\subsubsection*{Acknowledgments}
This work was supported by the MIUR grant PRIN 2017L5W2PT. A.G. thanks the University of Pisa for hospitality during the ongoing work.
We thank Daniele Gaggero and Paolo Panci for very useful discussions.

\end{document}